\documentclass[aps,prl,twocolumn,superscriptaddress,altaffilletter,nobibnotes,nofootinbib]{revtex4-1}
\usepackage{graphicx}  
\usepackage{dcolumn}   
\usepackage{bm}        
\usepackage{amssymb}   
\usepackage{lineno}
\usepackage{draftcopy}
\usepackage{hyperref}

\newcommand{\pt}{$p_{\rm{t}}$~}
\newcommand{\dca}{$dca$~}
\newcommand{\hijingb}{HIJING/B}

\newcommand{\ITS}{\rm{ITS}}
\newcommand{\SPD}{\rm{SPD}}
\newcommand{\SDD}{\rm{SDD}}
\newcommand{\SSD}{\rm{SSD}}
\newcommand{\TPC}{\rm{TPC}}

\newcommand{\pbarp}{$\overline{\rm p}/{\rm p}~$}

\usepackage{preprintcover}
\PreprintIdNumber{}
\PreprintCoverPaperTitle{Midrapidity antiproton-to-proton ratio in pp collisions at
$\sqrt{s} = 0.9$ and $7$~TeV measured by the ALICE experiment}
\PreprintCoverAbstract{The ratio of the yields of antiprotons to protons in pp collisions has been 
measured by the ALICE experiment at $\sqrt{s} = 0.9$ and $7$~TeV during the 
initial running periods of the Large Hadron Collider(LHC). The measurement 
covers the transverse momentum interval $0.45 < p_{\rm{t}} < 1.05$~GeV/$c$ 
and rapidity $|y| < 0.5$. The ratio is measured to be 
$R_{|y| < 0.5} = 0.957 \pm 0.006 (stat.) \pm 0.014 (syst.)$ 
at $0.9$~TeV and $R_{|y| < 0.5} = 0.991 \pm 0.005 (stat.) \pm 0.014 (syst.)$ 
at $7$~TeV and it is independent of both rapidity and transverse momentum. The 
results are consistent with the conventional model of baryon-number transport 
and set stringent limits on any additional contributions to baryon-number 
transfer over very large rapidity intervals in pp collisions.}
\PreprintJournalName{PRL}

\begin{document}


\title{Midrapidity antiproton-to-proton ratio in pp collisions at \\
$\sqrt{s} = 0.9$ and $7$~TeV measured by the ALICE experiment}

\collaboration{ALICE Collaboration}

\author{K.~Aamodt}
\affiliation{Department of Physics, University of Oslo, Oslo, Norway} 
\author{N.~Abel}
\affiliation{Kirchhoff-Institut f\"{u}r Physik, Ruprecht-Karls-Universit\"{a}t Heidelberg, Heidelberg, Germany} 
\author{U.~Abeysekara}
\affiliation{Physics Department, Creighton University, Omaha, NE, United States} 
\author{A.~Abrahantes~Quintana}
\affiliation{Centro de Aplicaciones Tecnol\'{o}gicas y Desarrollo Nuclear (CEADEN), Havana, Cuba} 
\author{A.~Abramyan}
\affiliation{Yerevan Physics Institute, Yerevan, Armenia} 
\author{D.~Adamov\'{a}}
\affiliation{Nuclear Physics Institute, Academy of Sciences of the Czech Republic, \v{R}e\v{z} u Prahy, Czech Republic} 
\author{M.M.~Aggarwal}
\affiliation{Physics Department, Panjab University, Chandigarh, India} 
\author{G.~Aglieri~Rinella}
\affiliation{European Organization for Nuclear Research (CERN), Geneva, Switzerland} 
\author{A.G.~Agocs}
\affiliation{KFKI Research Institute for Particle and Nuclear Physics, Hungarian Academy of Sciences, Budapest, Hungary} 
\author{S.~Aguilar~Salazar}
\affiliation{Instituto de F\'{\i}sica, Universidad Nacional Aut\'{o}noma de M\'{e}xico, Mexico City, Mexico} 
% ...

\vspace{0.1cm}

\date{\today}

\begin{abstract}

The ratio of the yields of antiprotons to protons in pp collisions has been 
measured by the ALICE experiment at $\sqrt{s} = 0.9$ and $7$~TeV during the 
initial running periods of the Large Hadron Collider(LHC). The measurement 
covers the transverse momentum interval $0.45 < p_{\rm{t}} < 1.05$~GeV/$c$ 
and rapidity $|y| < 0.5$. The ratio is measured to be 
$R_{|y| < 0.5} = 0.957 \pm 0.006 (stat.) \pm 0.014 (syst.)$ 
at $0.9$~TeV and $R_{|y| < 0.5} = 0.991 \pm 0.005 (stat.) \pm 0.014 (syst.)$ 
at $7$~TeV and it is independent of both rapidity and transverse momentum. The 
results are consistent with the conventional model of baryon-number transport 
and set stringent limits on any additional contributions to baryon-number 
transfer over very large rapidity intervals in pp collisions.
\vspace{1.5cm}
\end{abstract}


\maketitle
\linenumbers

In inelastic non-diffractive proton-proton collisions at very high
energy, the incoming projectile breaks up into several hadrons which
emerge after the collision in general under small angles along the
original beam direction. The deceleration  of the incoming proton,
or more precisely of the conserved baryon number associated with the
beam particles, is often called ``baryon-number transport'' and has been
debated theoretically for some time
\cite{Ref:RossiVeneziano,Ref:QGSM,Ref:StringJunction,Ref:Kopeliovich,Ref:Kharzeev,Ref:QGSMMerino,Ref:HijingB}.

One mechanism responsible for baryon-number transport is the break-up of 
the proton into a diquark--quark configuration~\cite{Ref:QGSM}. The diquark 
hadronizes after the reaction with some longitudinal momentum $p_z$ into a 
new particle, which carries the baryon number of the incoming proton. This 
baryon-number transport is usually quantified in terms of the rapidity loss
$\Delta y = y_{\rm{beam}} - y_{\rm{baryon}}$, where $y_{\rm{beam}}$
($y_{\rm{baryon}}$) is the rapidity of the incoming beam (outgoing 
baryon)\footnote{The rapidity $y$ is defined as 
$y = 0.5 \ln\left[\left(E+p_z\right)/\left(E-p_z\right)\right]$; 
rapidity $y = 0$ corresponds to longitudinal momentum $p_z = 0$ of the 
baryon in the center-of-mass system and 
$\Delta y = \ln{(\sqrt{s}/m_{\rm p})}$.}.

However, diquarks in general retain a large fraction of the proton momentum
and therefore stay close to beam rapidity, typically within one or two units.
Therefore, additional processes have been proposed to transport the baryon
number over larger distances in rapidity, in particular via purely gluonic
exchanges, where the proton breaks up into three quarks. The baryon number 
resides with a non-perturbative  configuration of gluon fields, the 
so-called ``baryon string junction'', which connects the valence quarks
\cite{Ref:RossiVeneziano,Ref:StringJunction}. In this picture, baryon-number 
transport is suppressed exponentially with the rapidity interval $\Delta y$, 
proportional to $\exp\left[\left(\alpha_{\rm{J}}-1\right)\Delta y\right]$, 
where $\alpha_{\rm{J}}$ is identified in the Regge model as the intercept of 
the trajectory for the corresponding exchange in the $t$-channel. If the 
string junction intercept is approximated with the one of the standard Reggeon 
(or meson), $\alpha_{\rm {J}} \approx 0.5$, baryon transport will approach 
zero with increasing $\Delta y$. If the intercept of the pure string junction 
is $\alpha_{\rm J} \approx 1$, as motivated by perturbative QCD 
\cite{Ref:Kopeliovich}, it will approach a constant and finite value.

The LHC, being by far the highest energy proton--proton collider, opens the
possibility to investigate baryon transport over very large rapidity intervals
by measuring the antiproton-to-proton production ratio at midrapidity,
$R = N_{\overline{\rm p}}/N_{\rm p}$, or equivalently, the proton--antiproton 
asymmetry, 
$A = (N_{\rm p} - N_{\overline{\rm p}}) / (N_{\rm p} + N_{\overline{\rm p}})$.
Most of the (anti)protons at midrapidity are created in baryon--antibaryon
pair production, implying equal yields. Any excess of protons over antiprotons
is therefore associated with the baryon-number transfer from the incoming
beam. Note that such a study has not been carried out in high-energy
proton--antiproton colliders (Sp$\overline{\rm p}$S, Tevatron) because of the 
symmetry of the initial system at midrapidity. Model predictions for the ratio 
$R$ at LHC energies range from unity, i.e., no baryon-number transfer to 
midrapidity, down to about 0.9 in models where the string junction transfer 
is not suppressed with the rapidity interval ($\alpha_{\rm {J}} \approx 1$).

In this letter, we describe the measurement of the \pbarp ratio at
midrapidity in non-diffractive pp collisions at center-of-mass energies
$\sqrt{s} = 0.9$~TeV and $7$~TeV ($\Delta y \approx 6.9$--$8.9$), with
the ALICE experiment at the LHC.

\vspace{0.1 cm}
ALICE, which is the dedicated heavy-ion detector at the LHC, consists of 18
detector sub-systems \cite{Ref:ALICE, Ref:ALICEPPR}. The central tracking
systems used in the present analysis are located inside a
solenoidal magnet (B = 0.5~T); they are optimized to provide
good momentum resolution and particle identification (PID) over a broad
momentum range, up to the highest multiplicities expected for heavy ion
collisions at the LHC. All detector systems were commissioned and aligned
during several months of cosmic-ray data-taking in 2008 and 2009
\cite{Ref:ALICECommissioning,Ref:ALICEITS}.

Collisions occur inside a beryllium vacuum pipe (3~cm in radius and 800~$\mu$m
thick) at the center of the ALICE detector. The tracking system in the ALICE
central barrel has full azimuth coverage within the pseudo-rapidity window
$|\eta| < 0.9$. The following detector sub-systems were used in this analysis:
the \textit{Inner Tracking System} (\ITS) \cite{Ref:ALICEITS}, the 
\textit{Time Projection Chamber} (\TPC) \cite{Ref:ALICETPC} and the VZERO 
detector \cite{Ref:ALICE}.

The \ITS~consists of six cylindrical layers of silicon detectors with radii 
of 3.9/7.6~cm (Silicon Pixel Detectors--\SPD), 15.0/23.9~cm (Silicon Drift 
Detectors--\SDD) and 38/43~cm (Silicon Strip Detectors--\SSD). They provide 
full azimuth coverage for tracks matching the acceptance of the \TPC~
($|\eta| < 0.9$). 

The \TPC~is the main tracking detector of the central barrel.  The detector
is cylindrical in shape with an active volume of inner radius 85~cm, outer
radius of 250~cm and an overall length along the beam direction of 500 cm.

Finally, the VZERO detector consists of two arrays of 32 scintillators each,
which are placed around the beam pipe on either side of the interaction
region) at $z = 3.3$~m and  $z = -0.9$~m, covering the pseudorapidity ranges
$2.8 < \eta < 5.1$ and  $-3.7 < \eta < -1.7$, respectively
\cite{Ref:ALICENchargedpapers}. A detailed description of the ALICE detectors,
its components, and their performance can be found in \cite{Ref:ALICE}.

\begin{figure}
\includegraphics[width=\linewidth]{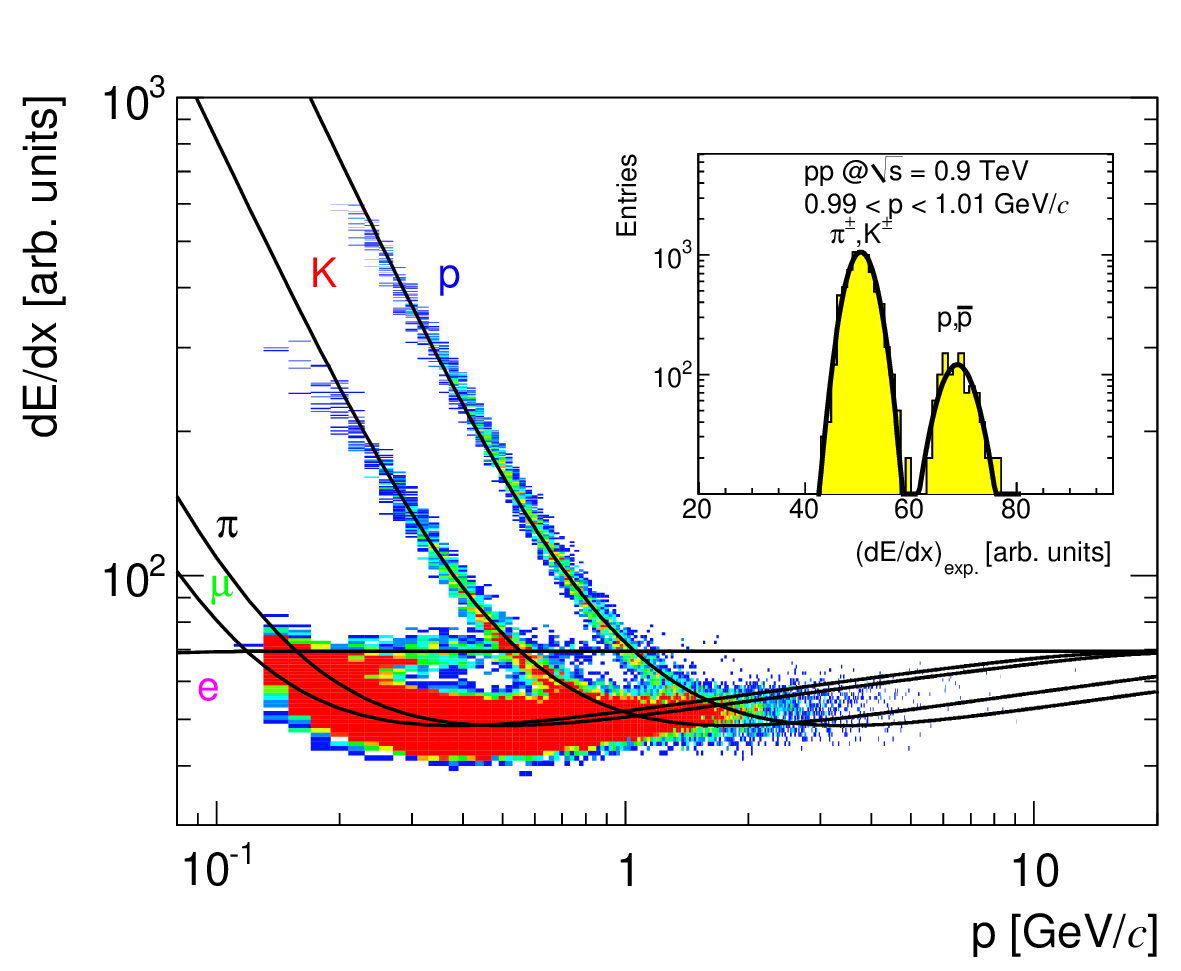}
\caption{(Color online) The measured ionization per unit length as a
function of particle momentum (both charges) in the \TPC~gas. The curves 
correspond to expected energy loss \cite{Ref:Aleph} for different particle 
types. The inset shows the measured ionization for tracks with 
$0.99 < p < 1.01$~GeV/$c$. The lines are Gaussian fits to the data.}
\label{fig:dEdxPTPC}
\end{figure}

Data from 2.8 ($\sqrt{s} = 0.9$~TeV) and 4.2 ($\sqrt{s} = 7$~TeV) million 
pp collisions, recorded during the first LHC runs (December~2009, 
March--April~2010) were used for this analysis. The events were recorded 
with both field polarities for each energy. The trigger required 
a hit in one of the VZERO counters or in the \SPD~detector, i.e., 
at least one charged particle anywhere in the 8 units of pseudorapidity 
covered by these trigger detectors \cite{Ref:ALICENchargedpapers}. In 
addition, the trigger required a coincidence between the signals from two 
beam pick-up counters, one on each side of the interaction region, 
indicating the presence of passing bunches.

Beam-induced background was reduced to a negligible level ($< 0.01\%$) with
the help of the timing information from the VZERO counters
\cite{Ref:ALICENchargedpapers} and by requiring a reconstructed primary
vertex (calculated from the \SPD) within $\pm 1$~cm perpendicular to and
$\pm 10$~cm along the beam axis.

Measurements of momentum and particle identification are performed using
information from the \TPC~detector, which measures the ionization in the
\TPC~gas and the particle trajectory with up to 159 space points. In order 
to ensure a good track quality, a minimum of 80 clusters was required per 
track in the \TPC~and at least two hits in the \ITS~of which at least one 
is in the \SPD. In order to reduce the contamination from background and 
secondary tracks (e.g.\,(anti)protons originating from weak hyperon decays 
or secondary interactions in the material), a cut was imposed on the 
distance of closest approach ($dca$) of the track to the primary vertex in 
the $xy$ (transverse) plane, which varied from 2.65 to 1.8~mm (2.33 to 
1.5~mm for the 7~TeV data) for the lowest ($0.45 < p_{\rm{t}} < 0.55$~GeV/$c$) 
and highest ($0.95 < p_{\rm{t}} < 1.05$~GeV/$c$) {\pt}~bins, respectively. 
This cut corresponds to $5\sigma$ of the measured \dca resolution for each 
momentum bin.

Particles are identified using their specific ionization
(${\rm d}E/{\rm d}x$) in the
\TPC~gas~\cite{Ref:ALICETPC}. Figure~\ref{fig:dEdxPTPC} shows the ionization
(truncated mean) as a function of particle momentum together with the
expected curves~\cite{Ref:Aleph} for different particle species. The inset
shows the measured ${\rm d}E/{\rm d}x$ for tracks in the momentum range
$0.99 < p < 1.01$~GeV/$c$  with clearly separated peaks for (anti)protons 
and lighter particles. The ${\rm d}E/{\rm d}x$ resolution of the \TPC~is 
$5$, depending slightly on the number of \TPC~clusters and the track 
inclination angle. For this analysis, (anti)protons were selected within 
a band of $\pm 3\sigma$ around the expected value.

\begin{figure}
\includegraphics[width=\linewidth]{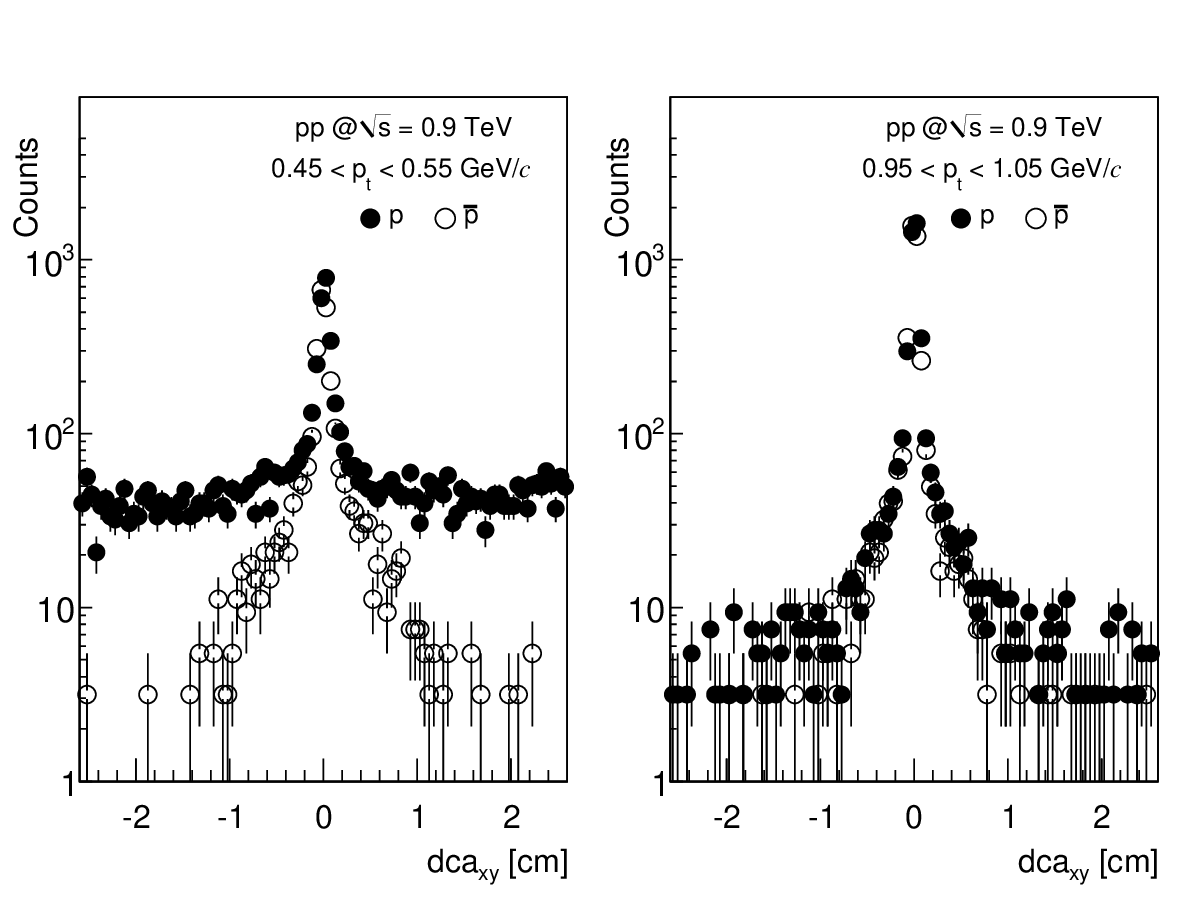}
\caption{The distance of closest approach ($dca$) distributions of $\rm{p}$
and $\overline{\rm{p}}$ for the lowest (left plot) and highest (right plot)
transverse momentum bins. The broad background of protons at low momentum
originates from secondary particles created in the detector material, whereas
the tails for both $\rm{p}$ and $\overline{\rm{p}}$ at high momentum (and for
$\overline{\rm{p}}$ at low momentum) arise from weak hyperon decays.}
\label{fig:correctionSecondaries}
\end{figure}

In order to assure uniform geometrical acceptance, high reconstruction
efficiency and unambiguous proton identification, we restrict the analysis
to protons and antiprotons in the rapidity range $|y| < 0.5$ and the momentum
range $0.45 < p < 1.05$~GeV/$c$.  The contamination of the proton
sample with electrons or pions and kaons is negligible ($< 0.1\%$) even at 
the highest momentum bins, and in addition essentially charge symmetric.


Most instrumental effects associated with the acceptance, reconstruction
efficiency, and resolution are identical for primary protons and anti-protons
and therefore cancel in the ratio. However, because of significant differences
in the relevant cross sections, anti-protons are more likely than protons to
be absorbed or elastically scattered\footnote{Particles undergoing elastic
scattering in the inner detectors can still be reconstructed in the TPC but
the corresponding ITS hits will in general not be associated to the track if
the scattering angle is large.} within the detector, and a non negligible
background in the proton sample arises from secondary interactions in the
beam pipe and inner layers of the detector.

In order to correct for the difference between p--A and 
$\overline{\rm{p}}$--A elastic and inelastic reactions in the 
detector material, detailed Monte Carlo simulations based on 
GEANT3 \cite{Ref:GEANT} and FLUKA \cite{Ref:FLUKA} were performed.
These corrections rely in particular on the proper description of 
the interaction cross sections used as input by the transport models. 
These values were therefore compared with experimental measurements
\cite{Ref:XSection,Ref:XSectionElastic}. While p--A cross sections 
are similar in both models and in agreement with existing data, 
GEANT3 (as well as the current version of GEANT4) significantly 
overestimates the measured inelastic cross sections for antiprotons in 
the relevant momentum range by about a factor of two, whereas FLUKA 
describes the data very well. Concerning elastic scattering, where only 
a limited data set is available for comparison, GEANT3 cross sections 
are about $25\%$ above FLUKA, the latter being again closer to the 
measurements. We therefore used the FLUKA results to account for the 
difference of $\rm{p}$ and $\overline{\rm{p}}$ cross sections, which 
amount to a correction of the \pbarp ratio by $8 \%$ and $3.5\%$ for 
absorption and elastic scattering, respectively.

The contamination of the proton sample due to secondaries originating from
interactions with the detector material was directly measured with the
data and subtracted. Most of these background tracks do not point back 
to the interaction vertex and can therefore be excluded with a \dca cut. 
Figure~\ref{fig:correctionSecondaries} shows the \dca distributions of 
$\rm{p}$ and $\overline{\rm{p}}$ for the lowest (left panel) and the 
highest (right panel) transverse momentum bins. Secondary protons are 
clearly visible in the left plot due to their wide \dca distribution. At 
higher momenta the background of secondary protons becomes very small. 
The remaining tails visible in the \dca distributions are due to 
(anti)protons originating from weak decays. The background of secondary 
protons, which remains after the \dca cut under the peak of primaries, 
is subtracted by determining its shape from Monte Carlo simulations and 
adjusting the amount to the data at large values of the $dca$. This 
correction is calculated and applied differentially as a function of $y$ 
and $p_{\rm{t}}$; it varies between $14\%$ for the lowest and less than 
$0.3\%$ for the highest transverse momentum bins.

The contamination coming from feed-down (i.e., (anti)protons originating
from the weak decay of $\Lambda$ and $\bar{\Lambda}$) was subtracted in a 
similar way by parametrization and fitting to the data of the respective 
simulated \dca distributions. This correction ranges from $20 \%$ 
to $12 \%$ for the lowest and highest \pt bins, respectively.

The main sources of systematic uncertainties are the detector material
budget, the (anti)proton reaction cross section, the subtraction of
secondary protons and the accuracy of the detector response simulations
(see Table~\ref{tab:systematic}). The amount of material in the central
part of ALICE is very low, corresponding to about $10 \%$ of
a radiation length on average between the vertex and the active volume
of the \TPC. It has been studied with collision data and adjusted in the
simulation based on the analysis of photon conversions. The current
simulation reproduces the amount and spatial distribution of reconstructed
conversion points in great detail, with a relative accuracy of a few percent.
Based on these studies, we assign a systematic uncertainty of
$7 \%$ to the material budget. By changing the material in the simulation
by this amount, we find a variation of the final ratio $R$ of less than
$0.5 \%$.

The experimentally measured $\overline{\rm{p}}$--$\rm{A}$ reaction cross
sections are determined with a typical accuracy better than $5\%$
\cite{Ref:XSection}. We assign a $10\%$ uncertainty to the absorption 
correction as calculated with FLUKA, which leads to a $0.8\%$ 
uncertainty in the ratio $R$.  By comparing GEANT3 with FLUKA and with
the experimentally measured elastic cross-sections, the corresponding
uncertainty was estimated to be $0.8\%$, which corresponds to the difference
between the correction factors calculated with the two models.

By changing the event selection, analysis cuts and track quality requirements
within reasonable ranges, we find a maximum deviation of the results of
$0.4 \%$, which we assign as systematic uncertainty to the accuracy of
the detector simulation and analysis corrections.

The uncertainty resulting from the subtraction of secondary protons and 
from the feed-down corrections was estimated to be $0.6 \%$ by using 
different functional forms for the background subtraction and for the 
contribution of the hyperon decay products. 

The contribution of diffractive reactions to our final event sample was
studied with different event generators and was found to be less than $3 \%$,
resulting into a negligible contribution ($< 0.1 \%$) to the systematic
uncertainty.

Finally, the complete analysis was repeated using only \TPC~information
(i.e., without using any of the \ITS~detectors). The resulting difference 
was negligible at both energies ($< 0.1 \%$).

\begin{table}[tb]
\caption{Systematic uncertainties of the \pbarp ratio.}
\centering
\begin{tabular}{cccc}
\hline
\hline
\multicolumn{4}{c}{Systematic Uncertainty} \\
\hline
\multicolumn{2}{l}{Material budget} & \multicolumn{2}{c}{$0.5 \%$} \\
\multicolumn{2}{l}{Absorption cross section} & \multicolumn{2}{c}{$0.8 \%$} \\
\multicolumn{2}{l}{Elastic cross section} & \multicolumn{2}{c}{$0.8 \%$} \\
\multicolumn{2}{l}{Analysis cuts} & \multicolumn{2}{c}{$0.4 \%$} \\
\multicolumn{2}{l}{Corrections (secondaries/feed-down)} & \multicolumn{2}{c}{$0.6 \%$} \\
\hline
\multicolumn{2}{l}{Total} & \multicolumn{2}{c}{$1.4 \%$} \\
\hline
\hline
\end{tabular}
\label{tab:systematic}
\end{table}

Table \ref{tab:systematic} summarizes the contribution to the systematic
uncertainty from all the different sources. The total systematic uncertainty
is identical for both energies and amounts to $1.4\%$.

\begin{figure}
\includegraphics[width=\linewidth]{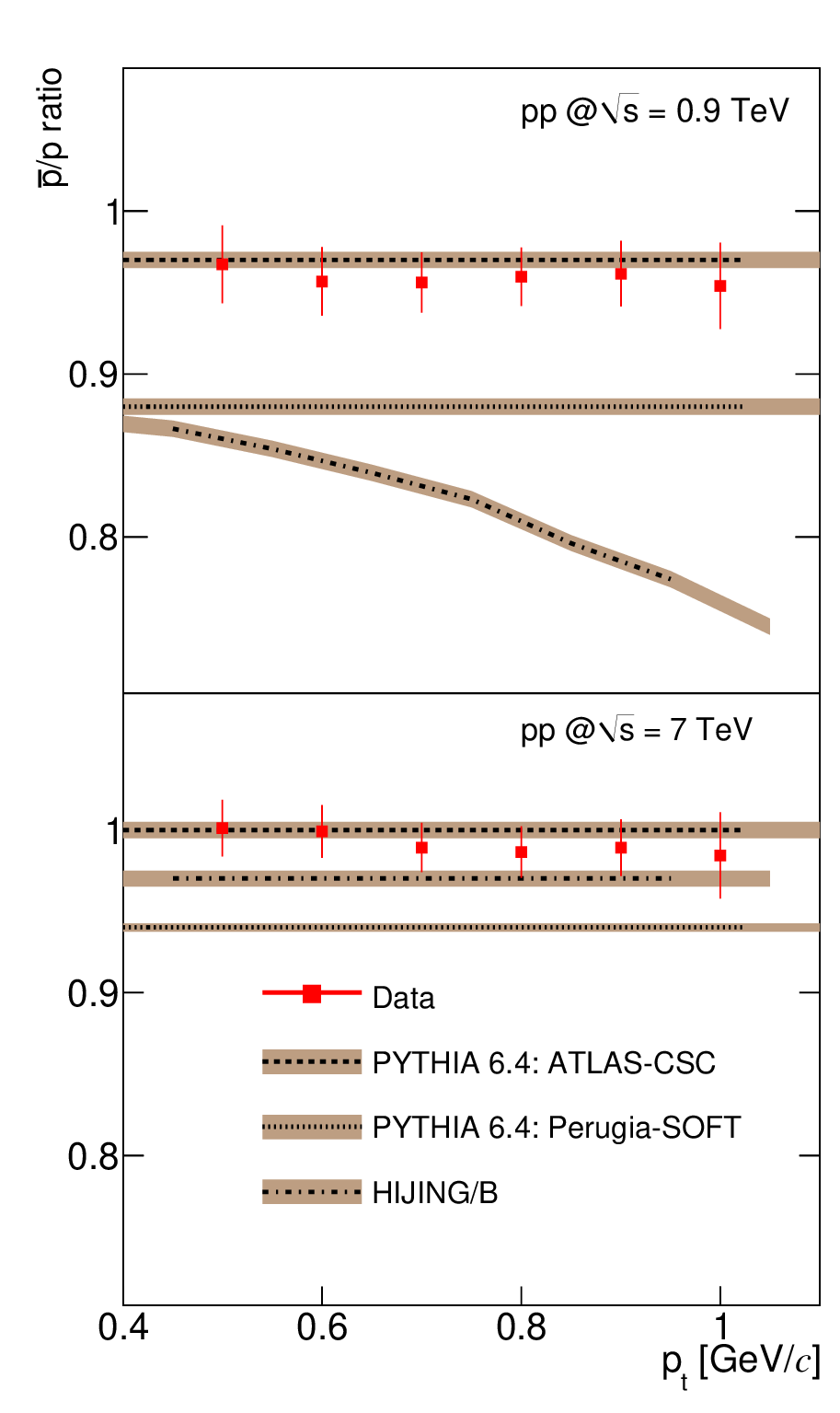}
\caption{(Color online) The \pt dependence of the \pbarp ratio integrated
over $|y| <0.5$ for pp collisions at $\sqrt{s} = 0.9$~TeV (top) and
$\sqrt{s} = 7$~TeV (bottom). Only statistical errors are shown for the data;
the width of the Monte Carlo bands indicates the statistical uncertainty of 
the simulation results.}
\label{fig:aliceratio}
\end{figure}

The final, feed-down corrected \pbarp ratio $R$ integrated within 
our rapidity and $p_{\rm{t}}$ acceptance rises from 
$R_{|y| < 0.5} = 0.957 \pm 0.006 (\emph{stat.}) \pm 0.014 (\emph{syst.})$ 
at $\sqrt{s} = 0.9$~TeV to 
$R_{|y| < 0.5} = 0.991 \pm 0.005 (\emph{stat.}) \pm 0.014 (\emph{syst.})$ 
at $\sqrt{s} = 7$~TeV. The difference in the \pbarp ratio, 
$0.034 \pm 0.008 (\emph{stat.})$, is significant because the 
systematic errors at both energies are fully correlated.

Within statistical errors, the measured ratio $R$ shows no dependence 
on transverse momentum (Fig.~\ref{fig:aliceratio}) or rapidity (data 
not shown). The ratio is also independent of momentum and rapidity 
for all generators in our acceptance, with the exception of \hijingb, 
which predicts a decrease with increasing transverse momentum for the 
lower energy.

\begin{table*}[tb]
\caption{The measured central rapidity \pbarp ratio compared to the 
predictions of different models (the statistical uncertainty in the models 
is less than $0.005$). The quoted errors for the ALICE points are the 
quadratic sum of statistical and systematic uncertainties.}
\centering
\begin{tabular}{llcc}
\hline
\hline
\multicolumn{2}{c}{Energy [TeV]} & 0.9 & 7 \\
\hline
\multicolumn{2}{l}{ALICE} & $0.957 \pm 0.015$ & $0.991 \pm 0.015$ \\
\hline
       & ATLAS-CSC Tune (306) & $0.96$ & $1.0$ \\
PYTHIA & Perugia-$0$ Tune (320) & $0.95$ & $1.0$ \\
       & Perugia-SOFT Tune (322) & $0.88$ & $0.94$ \\
\hline
     & $\epsilon = 0$ & $0.98$ & $1.0$ \\
QGSM & $\epsilon = 0.076$, $\alpha_{\rm{J}} = 0.5$ & $0.96$ & $0.99$ \\
     & $\epsilon = 0.024$, $\alpha_{\rm{J}} = 0.9$ & $0.89$ & $0.95$ \\
\hline
\multicolumn{2}{l}{\hijingb} & $0.83$ & $0.97$ \\
\hline
\hline
\end{tabular}
\label{tab:results}
\end{table*}

The data are compared with various model predictions for pp collisions
\cite{Ref:QGSMMerino,Ref:Pythia,Ref:HijingB} in Table~\ref{tab:results} 
(integrated values) and Fig.~\ref{fig:aliceratio}. The analytical QGSM 
model does not predict the \pt dependence and is therefore not included 
in Fig.~\ref{fig:aliceratio}. For both energies, two of the PYTHIA tunes 
\cite{Ref:Pythia} (ATLAS-CSC and Perugia-0) as well as the version of 
Quark--Gluon String Model (QGSM) with the value of the string junction 
intercept $\alpha_{\rm{J}} = 0.5$ \cite{Ref:QGSMMerino} describe the 
experimental values well, whereas QGSM without string junctions 
($\epsilon = 0$, $\epsilon$ is a parameter proportional to the probability 
of the string-junction exchange) is slightly above the data. 
\hijingb~\cite{Ref:HijingB}, unlike the above models, includes a particular 
implementation of gluonic string junctions to enhance baryon-number transfer. 
This model underestimates the experimental results, in particular at the 
lower LHC energy. Also, QGSM with a  value of the junction intercept 
$\alpha_{\rm{J}} = 0.9$ \cite{Ref:QGSMMerino} predicts a smaller ratio, as 
does the Perugia-SOFT tune of PYTHIA, which  also includes enhanced baryon 
transfer\footnote{We have checked that baryon transfer is the main reason 
for the different \pbarp ratios predicted by the models; the absolute yield 
of (anti)protons in our acceptance, which is dominated by pair production, 
is reproduced by the models to within $\pm 20 \%$.}.

\begin{figure}
\includegraphics[width=\linewidth]{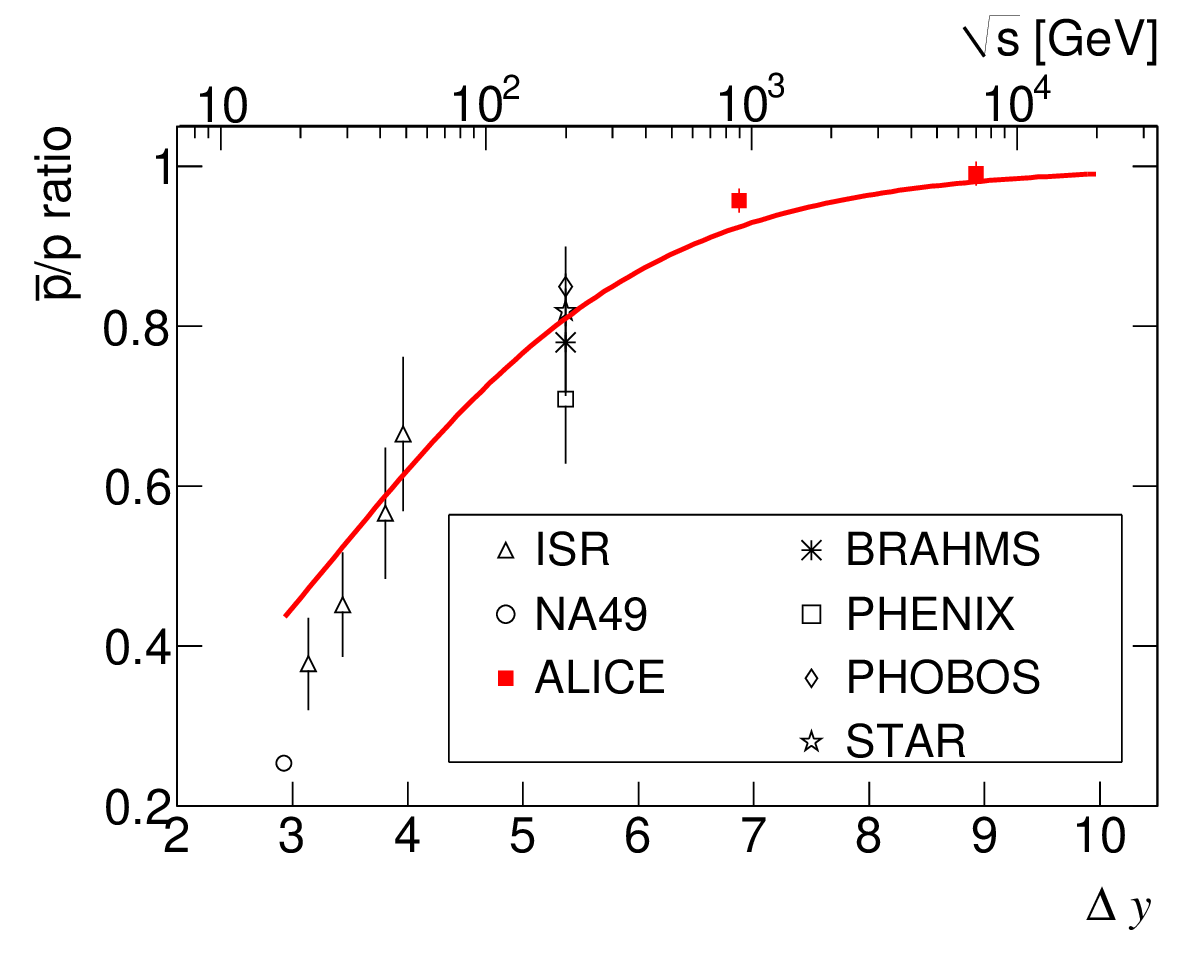}
\caption{(Color online) Central rapidity \pbarp ratio as a function of
the rapidity interval $\Delta y$ (lower axis) and center-of-mass energy 
(upper axis). Error bars correspond to the quadratic sum of statistical 
and systematic uncertainties for the RHIC and LHC measurements and to 
statistical errors otherwise.}
\label{fig:rapiditygapdependence}
\end{figure}

Figure~\ref{fig:rapiditygapdependence} shows a compilation of central 
rapidity measurements of the ratio $R$ in pp collisions as a function of 
center-of-mass energy (upper axis) and the rapidity interval $\Delta y$ 
(lower axis). The ALICE measurements correspond to $\Delta y = 6.87$ and 
$\Delta y = 8.92$ for the two energies, whereas the lower energy data 
points are taken from \cite{Ref:ExpSPS,Ref:ExpISR,Ref:ExpRHIC}. The \pbarp 
ratio rises from $0.25$ and $0.3$ at the SPS and  the lowest ISR energy, 
respectively, to a value of about $0.8$  at $\sqrt{s} = 200$~GeV, 
indicating that a substantial fraction of the baryon number associated with
the beam particles is transported over rapidity intervals of up to five units.

Although our measured midrapidity ratio $R$ at $\sqrt{s} = 0.9$~TeV is close
to unity, there is still a small but significant excess of protons over 
antiprotons corresponding to a $\rm{p}$--$\overline{\rm{p}}$ asymmetry of
$A = 0.022 \pm 0.003 (stat.) \pm 0.007 (syst.)$.
On the other hand, the ratio at $\sqrt{s} = 7$~TeV is consistent
with unity ($A = 0.005 \pm 0.003 (stat.) \pm 0.007 (syst.)$), which sets a
stringent limit on the amount of baryon transport over 9 units in rapidity.
The existence of a large value for the asymmetry even at infinite energy, 
which has been predicted to be $A = 0.035$ using 
$\alpha_{\rm J} = 1$~\cite{Ref:Kopeliovich}, is therefore excluded.

A rough approximation of the $\Delta y$ dependence of the ratio $R$ can 
be derived in the Regge model, where baryon pair production at very high 
energy is governed by Pomeron exchange and baryon transport by 
string-junction exchange~\cite{Ref:Kharzeev}. In this case the 
${\rm p}/\bar{\rm p}$ ratio  takes the simple form 
$1/R = 1 + C \exp[(\alpha_{\rm J} - \alpha_{\rm P}) \Delta y]$. We have 
fitted such a function to the data, using as value for the Pomeron intercept 
$\alpha_{\rm P} = 1.2$~\cite{Ref:alpha1.2} and $\alpha_{\rm J} = 0.5$, 
whereas $C$, which determines the relative contributions of the two diagrams, 
is adjusted to the measurements from ISR, RHIC, and LHC. The fit, shown in 
Fig.~\ref{fig:rapiditygapdependence}, gives a reasonable description of the 
data with only one free parameter ($C$), except at lower energies, where 
contributions of other diagrams cannot be neglected~\cite{Ref:Kharzeev}.
Adding a second string junction diagram with a larger 
intercept~\cite{Ref:Kopeliovich}, i.e.,
$1/R = 1 + C \exp[(\alpha_{\rm J} - \alpha_{\rm P}) \Delta y] + C' \exp[(\alpha_{\rm J'} - \alpha_{\rm P}) \Delta y]$ 
with $\alpha_{\rm J'} = 1$, does not improve the quality of the fit and its 
contribution is compatible with zero 
($C \approx 10$, $C' \approx -0.1 \pm 0.1$). In a similar spirit, our data 
could also be used to constrain other Regge-model inspired descriptions of 
baryon asymmetry, for example when the string-junction exchange is replaced 
by the ``odderon'', which is the analogue of the Pomeron with odd C-parity; 
see~\cite{Ref:QGSMMerino}.

In summary, we have measured the ratio of antiproton to proton 
production in the ALICE experiment at the CERN LHC collider at 
$\sqrt{s} = 0.9$ and $\sqrt{s} = 7$~TeV. Within our acceptance 
region ($|y| < 0.5$, $0.45 < p_{\rm{t}} < 1.05$~GeV/$c$), the 
ratio of antiproton-to-proton yields rises from 
$R_{|y| < 0.5} = 0.957 \pm 0.006 (stat.) \pm 0.014 (syst.)$
at 0.9 to a value close to unity 
$R_{|y| < 0.5} = 0.991 \pm 0.005 (stat.) \pm 0.014 (syst.)$
at 7~TeV. The \pbarp ratio is independent of both rapidity and 
transverse momentum. These results are consistent with standard 
models of baryon-number transport and set tight limits on any 
additional contributions to baryon-number transfer over very 
large rapidity intervals in pp collisions.


\section*{Acknowledgements}
\small{
We would like to thank Paola Sala, Alfredo Ferrari, Dmitri Kharzeev, Carlos Merino, Torbj\"{o}rn Sj\"{o}strand and Peter Skands for numerous and fruitful discussions on different topics of this paper. 

The ALICE collaboration would like to thank all its engineers and technicians for their invaluable contributions to the construction of the experiment and the CERN accelerator teams for the outstanding performance of the LHC complex.

The ALICE collaboration acknowledges the following funding agencies for their support in building and
running the ALICE detector:
Calouste Gulbenkian Foundation from Lisbon and Swiss Fonds Kidagan, Armenia;
Conselho Nacional de Desenvolvimento Cient\'{\i}fico e Tecnol\'{o}gico (CNPq), Financiadora de Estudos e Projetos (FINEP),
Funda\c{c}\~{a}o de Amparo \`{a} Pesquisa do Estado de S\~{a}o Paulo (FAPESP);
National Natural Science Foundation of China (NSFC), the Chinese Ministry of Education (CMOE)
and the Ministry of Science and Technology of China (MSTC);
Ministry of Education and Youth of the Czech Republic;
Danish Natural Science Research Council, the Carlsberg Foundation and the Danish National Research Foundation;
The European Research Council under the European Community's Seventh Framework Programme;
Helsinki Institute of Physics and the Academy of Finland;
French CNRS-IN2P3, the `Region Pays de Loire', `Region Alsace', `Region Auvergne' and CEA, France;
German BMBF and the Helmholtz Association;
Hungarian OTKA and National Office for Research and Technology (NKTH);
Department of Atomic Energy and Department of Science and Technology of the Government of India;
Istituto Nazionale di Fisica Nucleare (INFN) of Italy;
MEXT Grant-in-Aid for Specially Promoted Research, Ja\-pan;
Joint Institute for Nuclear Research, Dubna;
Korea Foundation for International Cooperation of Science and Technology (KICOS);
CONACYT, DGAPA, M\'{e}xico, ALFA-EC and the HELEN Program (High-Energy physics Latin-American--European Network);
Stichting voor Fundamenteel Onderzoek der Materie (FOM) and the Nederlandse Organisatie voor Wetenschappelijk Onderzoek (NWO), Netherlands;
Research Council of Norway (NFR);
Polish Ministry of Science and Higher Education;
National Authority for Scientific Research - NASR (Autontatea Nationala pentru Cercetare Stiintifica - ANCS);
Federal Agency of Science of the Ministry of Education and Science of Russian Federation, International Science and
Technology Center, Russian Academy of Sciences, Russian Federal Agency of Atomic Energy, Russian Federal Agency for Science and Innovations and CERN-INTAS;
Ministry of Education of Slovakia;
CIEMAT, EELA, Ministerio de Educaci\'{o}n y Ciencia of Spain, Xunta de Galicia (Conseller\'{\i}a de Educaci\'{o}n),
CEA\-DEN, Cubaenerg\'{\i}a, Cuba, and IAEA (International Atomic Energy Agency);
Swedish Reseach Council (VR) and Knut $\&$ Alice Wallenberg Foundation (KAW);
Ukraine Ministry of Education and Science;
United Kingdom Science and Technology Facilities Council (STFC);
The United States Department of Energy, the United States National
Science Foundation, the State of Texas, and the State of Ohio.}


\end{document}